\documentclass[12pt]{article}
\usepackage{amsmath}
\usepackage{graphicx,psfrag,epsf}
\usepackage{enumerate}
\usepackage{subfigure}
\usepackage{booktabs} 
\usepackage[round, sort]{natbib}
\setcitestyle{aysep={}}
\usepackage{hyperref}  
\hypersetup{hypertexnames=false,colorlinks=true,allcolors=,urlcolor=magenta }
\usepackage[capitalise]{cleveref}

\begin{document}


  \title{\bf Sounding Spider: An Efficient Way for Representing Uncertainties in High Dimensions}
  \author{Pamphile T. Roy, Sophie Ricci,  B\'en\'edicte Cuenot and Jean-Christophe Jouhaud\thanks{The authors acknowledge Bertrand Iooss and Michael Baudin from MRI (EDF R\&D) for helpful discussions on uncertainty quantification, HDR boxplot and support on OpenTURNS.
    The financial support provided by all the CERFACS shareholders (AIRBUS Group, Cnes, EDF, M\'et\'eo-France, ONERA, SAFRAN and TOTAL) is greatly appreciated and we thank them for enabling the achievement of such research activities.}\hspace{.2cm}\\
    CERFACS, France}
  \maketitle

\bigskip
\begin{abstract}
This article proposes a visualization method for multidimensional data based on: \emph{(i)} Animated functional Hypothetical Outcome Plots (f-HOPs); \emph{(ii)} 3-dimensional Kiviat plot; and \emph{(iii)} data sonification. In an Uncertainty Quantification (UQ) framework, such analysis coupled with standard statistical analysis tools such as Probability Density Functions (PDF) can be used to augment the understanding of how the uncertainties in the numerical code inputs translate into uncertainties in the quantity of interest (QoI).

In contrast with static representation of most advanced techniques such as functional Highest Density Region (HDR) boxplot or functional boxplot, f-HOPs is a dynamic visualization that enables the practitioners to infer the dynamics of the physics and enables to see functional correlations that may exist. While this technique only allows to represent the QoI, we propose a 3-dimensional version of the Kiviat plot to encode all input parameters. This new visualization takes advantage of information from f-HOPs through data sonification. All in all, this allows to analyse large datasets within a high-dimensional parameter space and a functional QoI in the same canvas. The proposed method is assessed and showed its benefits on two related environmental datasets.
\end{abstract}

\noindent%
\vfill

\newpage
\section{Introduction}
\label{sec:intro}

\emph{In vivo} and \emph{in silico} experiments are standard tools for reducing lag between design and industrialization phases~\citep{Montomoli2015}. These experiments make the most of cutting-edge instruments and High Performance Computing (HPC) resources to perform Uncertainty Quantification (UQ)~\citep{Sacks1989} studies dedicated for instance to parametric studies and design optimization. A system is hereafter described in terms of input and output spaces of potentially large dimension, related by a numerical solver for \emph{in silico} experiments. Generally speaking, UQ stands in: \emph{(i)} describing the uncertainties in the input space, \emph{(ii)} propagating these uncertainties onto the output space where the Quantity of Interest (QoI) is defined, \emph{(iii)} estimating and ranking the contribution of each source of uncertainty on the QoI. These contributions are estimated with sensitivity indices such as \emph{Sobol'} indices~\citep{Saltelli2007} that are based on variance decomposition. The uncertainty in the input space can then be reduced using optimization or data assimilation algorithms~\citep{AshBocquetNodet2017} when observations are available in order to reduce the uncertainty in the output. The ultimate goal is to assess a risk, for instance of failure or of exceeding a threshold.

\begin{figure}[!h]
\centering
\includegraphics[width=0.8\linewidth,keepaspectratio]{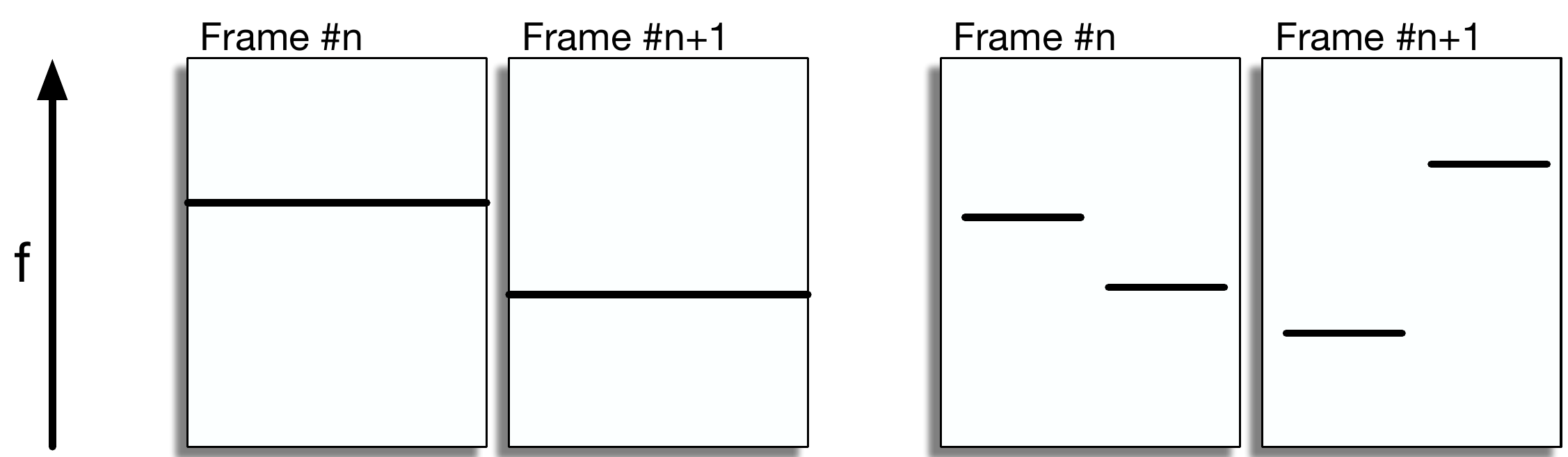}
\caption{Schematic representation of the HOPs method for single scalar output (\emph{left}) or multiple scalar outputs (\emph{right}).}
\label{fig:pattern_hop}
\end{figure}

In spite of a large literature on UQ, the community has yet to propose efficient ways to visualize uncertainties. Indeed, to the authors’ knowledge, there is no chapter dedicated to visualization in UQ reference books~\citep{Saltelli2007,Sullivan2015,HandbookUQ}. This remains to be investigated, especially for Computational Fluid Dynamics (CFD) applications~\citep{Moreland2016} that involve complex fields of large dimensions. Classical ways of visualizing standard statistics can lead to misinterpretation~\citep{Anscombe1973}. The challenge for state-of-the-art visualization solutions stands in the dimension of the data. Assuming that the dimension of the data is limited (a set of scalars), for instance when dealing with input data , canonical subplots of subspaces are adapted. Parallel coordinate plots~\citep{Inselberg1985} or Kiviat plot (or spider plot)~\citep{Hackstadt1994} that are represented in~\Cref{fig:sketch_Kiviat-parallel} offer an interesting alternative and share the same idea of dedicating one input variable (noted $x_i$) per axis; Kiviat (right panel) plot being the equivalent to parallel coordinates (left panel) plot in polar coordinates. 
\begin{figure}[!h]
\centering
\includegraphics[width=0.8\linewidth,keepaspectratio]{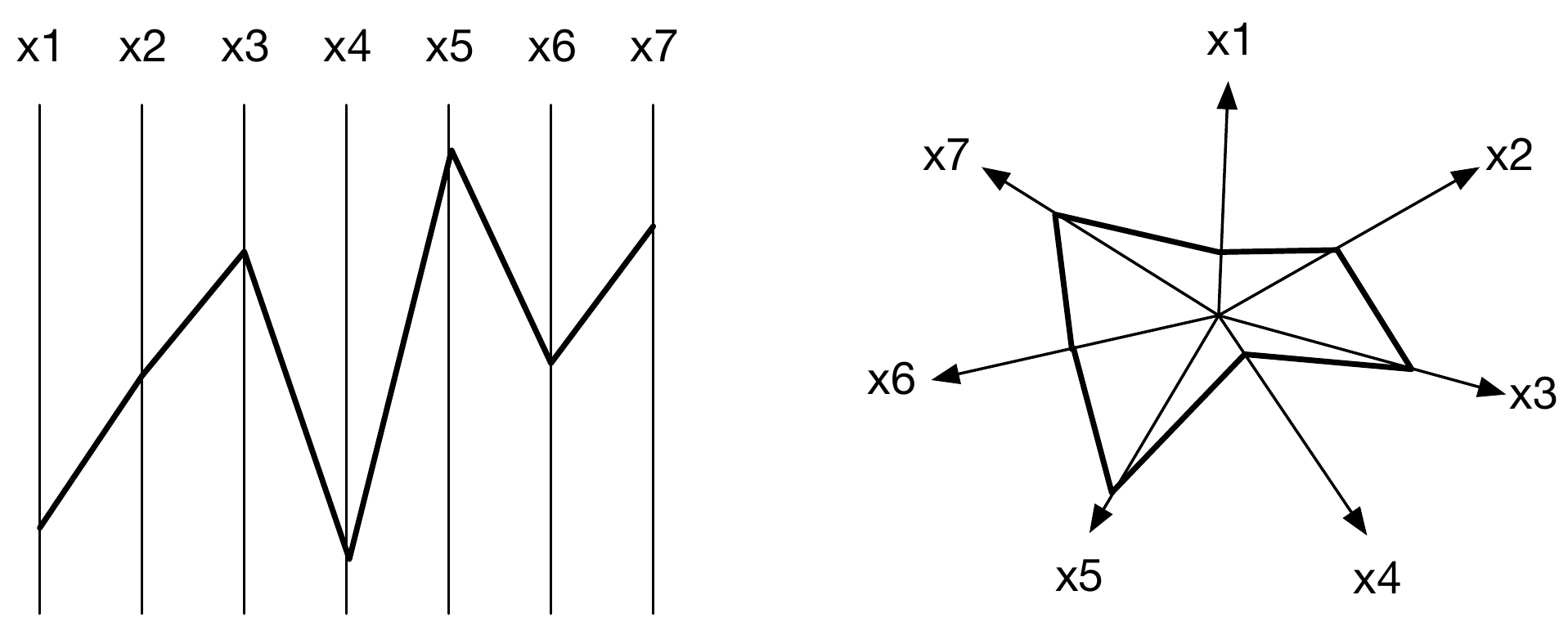}
\caption{Schematic representation of a parallel coordinate plot (\emph{left}) and a Kiviat plot (\emph{right}).}
\label{fig:sketch_Kiviatparallel}
\end{figure}
When the dimension of the data increases, for instance when dealing with \emph{functional} output fields discretized in both space and time on fine meshes, advanced strategies should be proposed. Different strategies are found in the literature to visualize statistics on the QoI~\citep{Potter2012a,Brodlie2012,Bonneau2014}. Beyond deterministic simulation, moving on to ensemble-based approaches, the dimension of the data further increases. A first approach is to look at each realization individually. For scalar QoI, a simple answer is found with the Hypothetical Outcome Plots (HOPs) technique~\citep{Ehlschlaeger1997} generalized by~\citep{Hullman2015} for a set of scalars as presented in~\cref{fig:pattern_hop}. HOPs consists in animating a succession of possible outcomes sampled from the data. It was proven to enable the exploration and the understanding of the dataset general characteristic, even for people lacking statistical background~\citep{Belia2005} with the possibility to visually grasp correlations in the outputs.

\begin{figure}[!h]
\centering
\includegraphics[width=\linewidth,keepaspectratio]{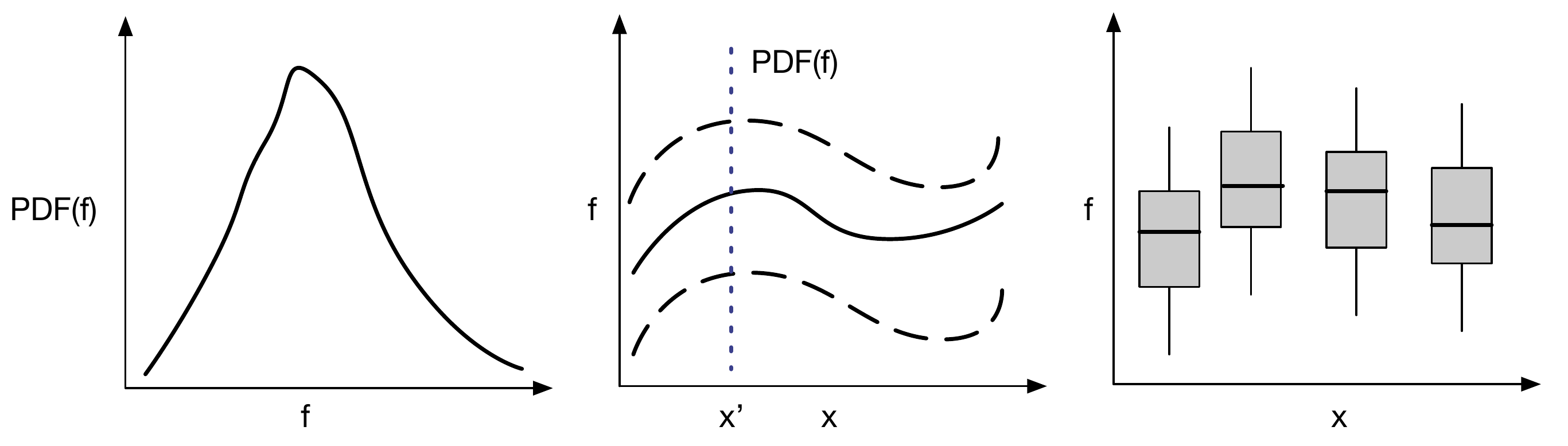}
\caption{Schematic representation of the PDF for a scalar output (\emph{left}), for a functional output (\emph{middle}) and with a boxplot solution (\emph{right}). The PDF mode is represented with a solid line, the standard deviation added/removed to/from the mean is represented with dashed lines.}
\label{fig:pattern_pdf}
\end{figure}

Statistical moments and Probability Density Function (PDF) plot are common tool for visualizing uncertainties in the context of ensemble simulation as illustrated in~\cref{fig:pattern_pdf}. They are useful for risk analysis as the probability of exceeding a threshold is directly observable. \Cref{fig:pattern_pdf}(a) displays the PDF for a scalar QoI (noted $f$) where the mean, the mode and extreme probabilities can be observed. \Cref{fig:pattern_pdf}(b) displays the principal mode of the PDF (solid black line) for a functional QoI discretized in the $x-$direction. The PDF standard deviation (added/removed to/from the mean) is plotted in dashed lines. These curves are computed for each $x$ independently and they do not represent a possible solution of the numerical solver. The box plot in \Cref{fig:pattern_pdf}(c) provides similar information while diminishing the false illusion of median/mean curves. It should be noted that users usually have a better understanding of frequency~\citep{Gigerenzer1995} than of PDFs and that there is a  general misinterpretation of confidence intervals~\citep{Belia2005}.

\begin{figure}[!h]
\centering
\includegraphics[width=0.6\linewidth,keepaspectratio]{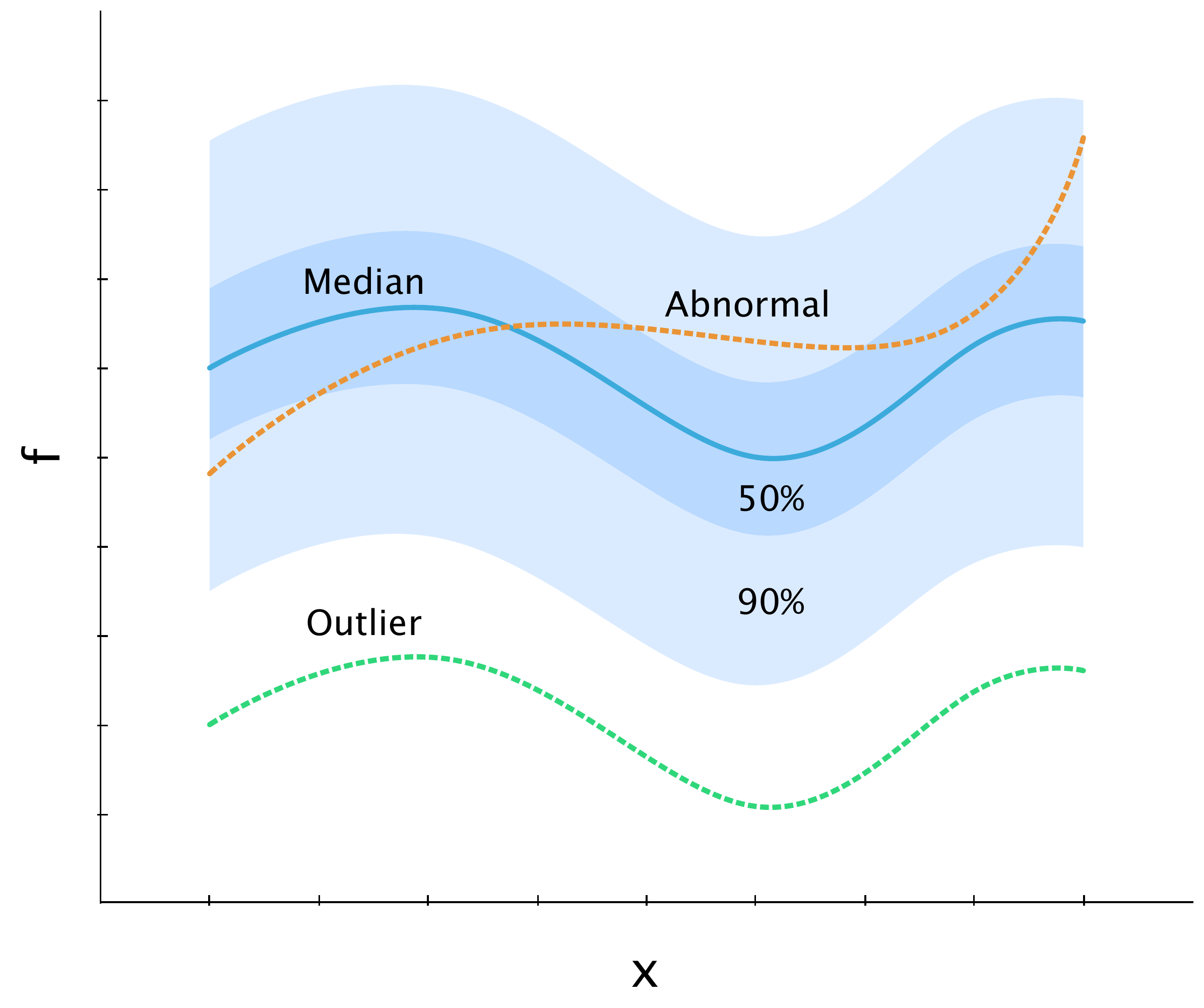}
\caption{Schematic representation of data set functional realizations characterized with HDR metrics. The median is represented by the solid blue line, the abnormal and outliers are represented by the red and green lines and the 50\% and 90\% quantiles are represented by blue shaded envelopes.}
\label{fig:pattern_hdr}
\end{figure}

A complementary approach based on density criteria was proposed by~\citep{Hyndman2009,Sun2011}; it is noted HDR for Highest Density Region and represented in~\cref{fig:pattern_hdr}. It allows to depict some statistics (for instance median or outliers) taking into account the functional QoI as a whole and working in a reduced space spanned by the most significant directions of the output space. Within this reduced space, metrics for functional outputs are computed such as the distance to the median (blue curve) so that abnormal or outlier outputs (red and green curves) can be detected and quantiles can be estimated (blue envelope). Each curve represents either a realization within the data set or an additional realization sampled in the reduced space; thus functional characteristics such as spatial or temporal correlation are preserved.  From~\citep{Popelin2013,Ribes2015}, the HDR method is more robust to outlier detection than other methods such as functional boxplot~\citep{Sun2011}.

This article proposes a solution to visualize high input and output dimensions. This solution relies on HDR metrics and a 3-dimensional version of Kiviat plot. For the 3D-Kiviat, each layer stands for a hypothetical outcome (a realization) coloured by a scalar value related to the QoI. This value is either the QoI at a given point and time or some distance computed with the HDR metrics. This solution can be augmented with data sonification to traduce the HDR metrics.

The paper is tailored as follows: \cref{sec:MatMet} presents existing material and methods for our study. \cref{sec:dataset} presents the dataset used for illustration purpose and \cref{sec:HDR} presents the HDR metrics. \Cref{sec:uqvisu} presents our innovative technique to visualize output uncertainties based on f-HOPS with sonification for HDR metrics. \Cref{sec:input} extends this advanced solution to both input and output uncertainties visualization building on 3D-Kiviat representation augmented with HDR metrics and sonification. Conclusion and discussion are finally given in \Cref{sec:ccl}.

\section{Materials and methods}
\label{sec:MatMet}
\subsection{Datasets}
\label{sec:dataset}

Visualization solutions proposed in this paper are applied on two datasets described in~\cref{tab:dataset} in order to state the applicability to various fields of applications. The first dataset has no input-output relation and only features a temporal output. The second dataset features an input-output relation with spatially varying output. The datasets are as follows:
\begin{itemize}
\item The \emph{El Ni\~{n}o} dataset is a well-known functional dataset~\citep{Hyndman2009}. It consists in a time series of monthly averaged Sea Surface Temperature (SST) in degrees Celsius spatially averaged over the Pacific Ocean region (0-10°S and 90-80°W) from January 1950 to December 2007. The QoI is a vector of size 12 and the data set gathers 58 realizations. Data originate from NOAA ERSSTv5's database available at\\ \href{http://www.cpc.ncep.noaa.gov/data/indices}{http://www.cpc.ncep.noaa.gov/data/indices}.
\item The \emph{Hydrodynamics} dataset gathers water levels (in m) computed with the 1-dimensional Shallow Water Equation MASCARET solver (opentelemac.org) for a 50~km reach of the Garonne river in South-West of France~\citep{Roy2017}. Uncertain inputs relate to 4 scalars: the friction coefficients of the river bed $Ks1, Ks2, Ks3$ defined over three homogeneous spatial areas, and the upstream boundary condition described by a constant scalar value for the inflow $Q$ in stationary flow. The QoI is a vector of size 463 (number of computation nodes for the 1D mesh) and an ensemble of 200 realizations is considered here. 
\end{itemize}

\begin{table}[!h]
\centering
\begin{tabular}{lccc}
\toprule
Dataset & Scalar input & Functional output & Sample size\\
\midrule 
El Ni\~no & - & 12 & 58\\
Hydrodynamics & 4 & 463 & 200\\
\bottomrule
\end{tabular}
\caption{Description of the El Ni\~no and Hydrodynamics datasets.}
\label{tab:dataset}
\end{table}

\subsection{Highest Density Region}
\label{sec:HDR}

The dataset output is considered as a matrix where each line corresponds to a realization. This matrix is decomposed by principal component analysis (PCA)~\citep{AnindyaChatterjee2000}. The modes are ordered by decreasing importance in terms of contribution to the variance and only a finite number of modes are kept. In this reduced space, the functional dataset of large dimensions is conveniently represented by a limited number of scalars mapped onto most significant directions that maximizes the variance of the QoI.
Within this reduced space, the median realization corresponds to the HDR location. The distance to this point is computed in the modal space; the farther a point is from the HDR, the less probable is the realization.

A multivariate Kernel Density Estimation (KDE)~\citep{Wand1995} technique is used to estimate the Probability Density Function (PDF) $\hat{f}(\mathbf{x_r})$ of this multivariate space. The density estimator is given by
\begin{align}
\hat{f}(\mathbf{x_r})&= \frac{1}{N_{s}}\sum_{i=1}^{N_{s}} K_{h_i}(\mathbf{x_r}-\mathbf{x_r}_i),
\end{align}
\noindent with $h_{i}$ the bandwidth for the \emph{i}th component and $K_{h_i}(.) = K(./h_i)/h_i$ the kernel chosen as a modal probability density centred and symmetric function. $K$ is the Gaussian kernel and $h_{i}$ are optimized given the data. From this KDE, the HDR reads
\begin{align}
R_\alpha = {x_r: \hat{f}(\mathbf{x_r}) \geq f_{\alpha}},
\end{align}
\noindent with $f_{\alpha}$ such that $\int_{R_\alpha} \hat{f}(\mathbf{x_r}) d x_r = 1 - \alpha$. The region has a probability of containing $1-\alpha$ points that verify the highest density estimate. Thus the median curve is defined as the inverse transform---from the reduced space into the original space---of $\arg \sup \hat{f}(\mathbf{x_r})$. Also 50\% and 90\% highest density regions are computed---with respectively $\alpha=0.5$ and $\alpha=0.1$.

\begin{figure*}[!h]               
\centering
\subfigure[]{
\includegraphics[width=0.45\linewidth,height=\textheight,keepaspectratio]{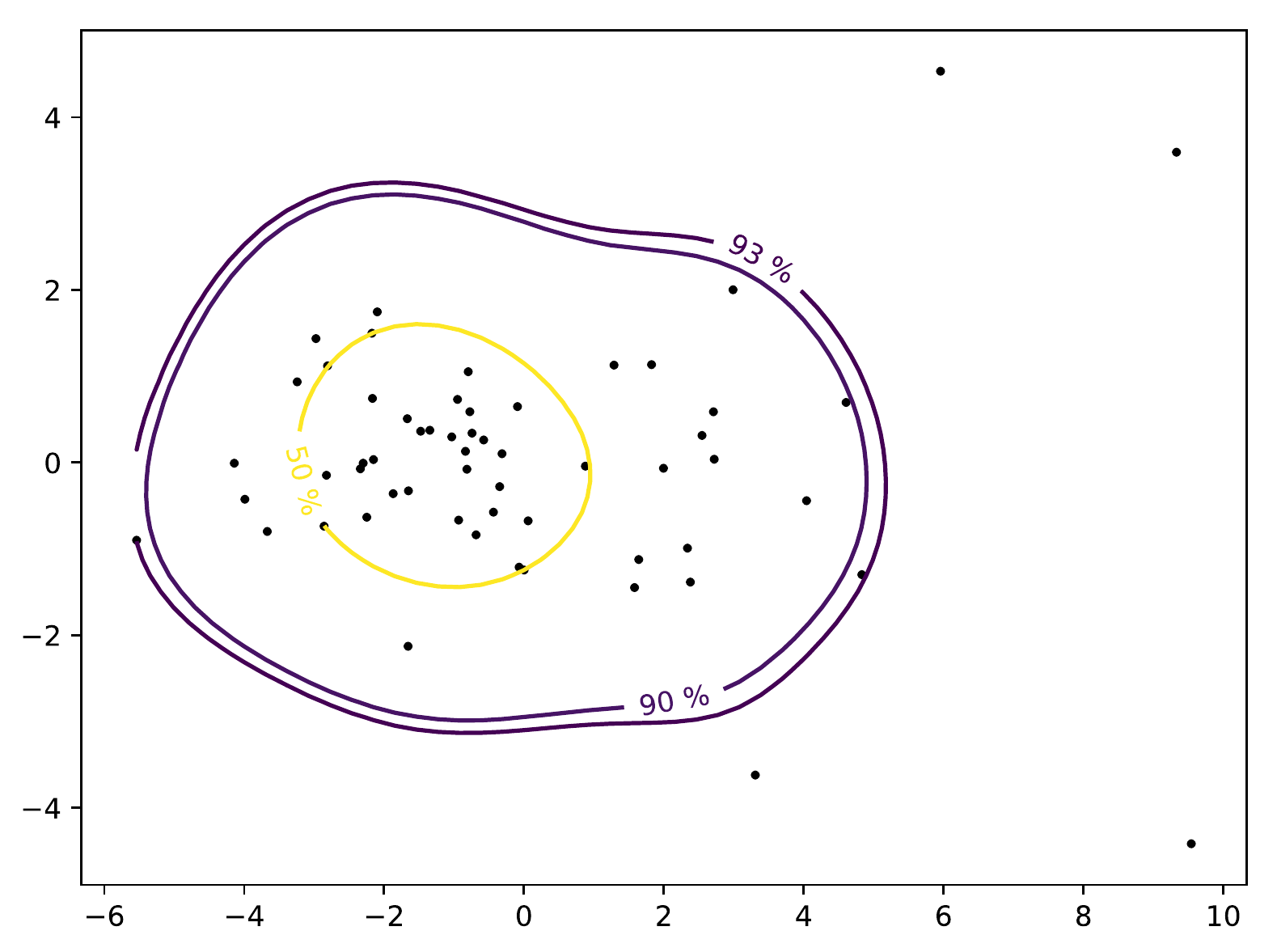}}
\subfigure[]{
\includegraphics[width=0.45\linewidth,height=\textheight,keepaspectratio]{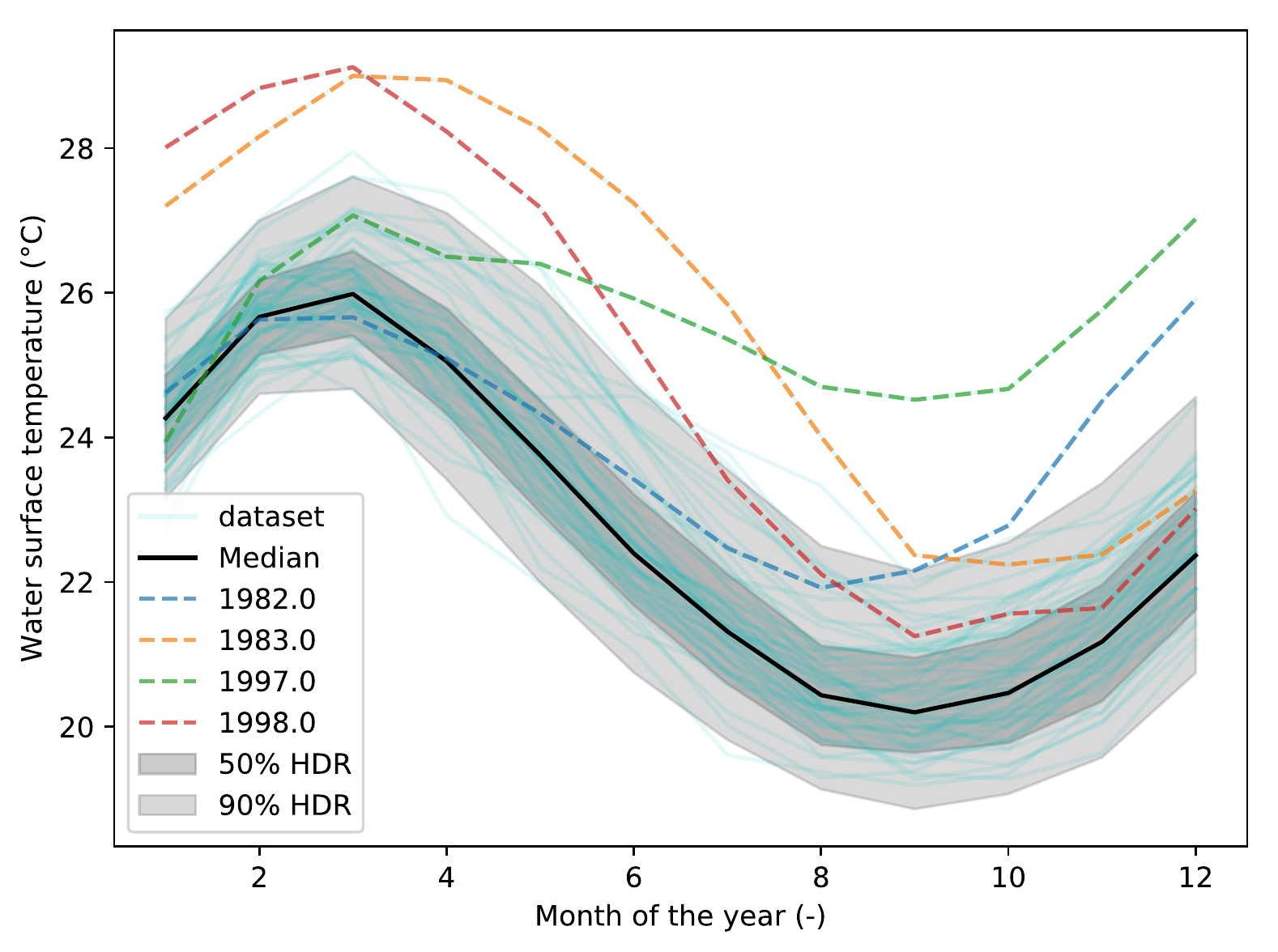}} 
\caption{HDR boxplot on the El Ni\~no dataset. \textbf{a} scatter plot of the 2-dimensional reduced space with each dot as a realization. \textbf{b} dataset visualization with each curve as a realization from the database. Shaded areas are confidence intervals, \emph{thick solid black} line is the mean realization and \emph{highlighted-dashed} curves are outliers.}
\label{fig:elnino}
\end{figure*}

\cref{fig:elnino} illustrates the HDR boxplot for the El Ni\~no dataset in the reduced space (left panel) when only two modes are retained ensuring that at least 80\% of the QoI variance is conserved and in the output space (right) panel. Each realization is characterized with respect to the HDR metric. In the modal space, each dot represents a realization within the dataset and the contouring represents the 50\% and 90\% quantiles. In the QoI physical space, cyan curves represent the realizations from the dataset, the outliers are coloured-dashed curves, the thick black curve is the median and the grey shaded areas represent 50\% and 90\% quantiles envelopes. It should be noted that additional realizations with chosen characteristics on the outputs could be drawn by sampling the input for specific HDR criteria.

\section{Uncertainty Visualization of Functional Output Data}
\label{sec:uqvisu}

\subsection{Dynamic Visualization of Functional Outputs' Statistics}
\label{subsec:fhop}

%



\begin{figure*}[!h]               
\centering
\subfigure[Frame \#$n$]{
\includegraphics[width=0.45\linewidth,height=\textheight,keepaspectratio]{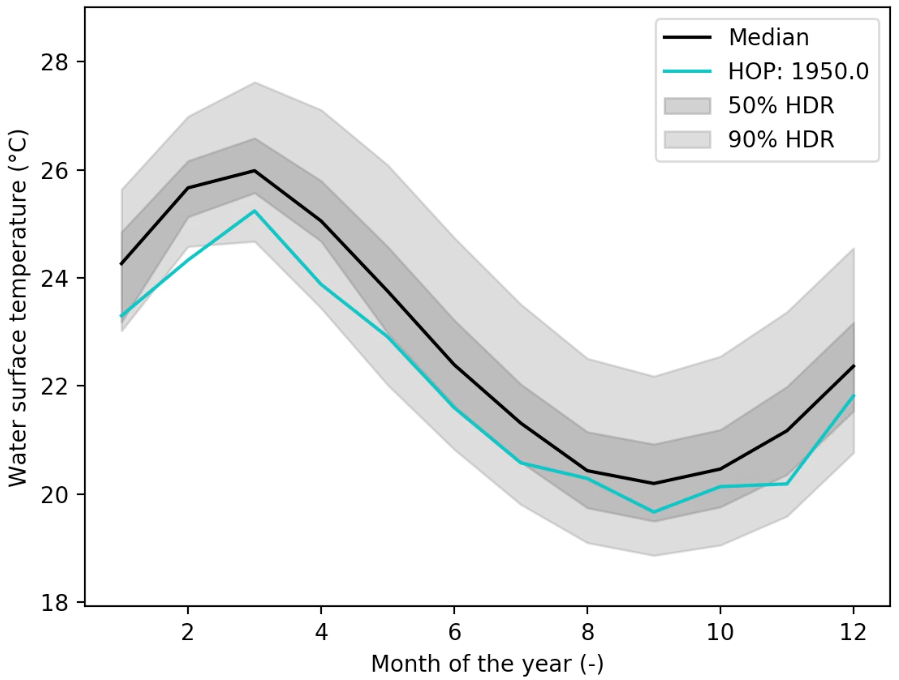}}
\subfigure[Frame \#$n+1$]{
\includegraphics[width=0.45\linewidth,height=\textheight,keepaspectratio]{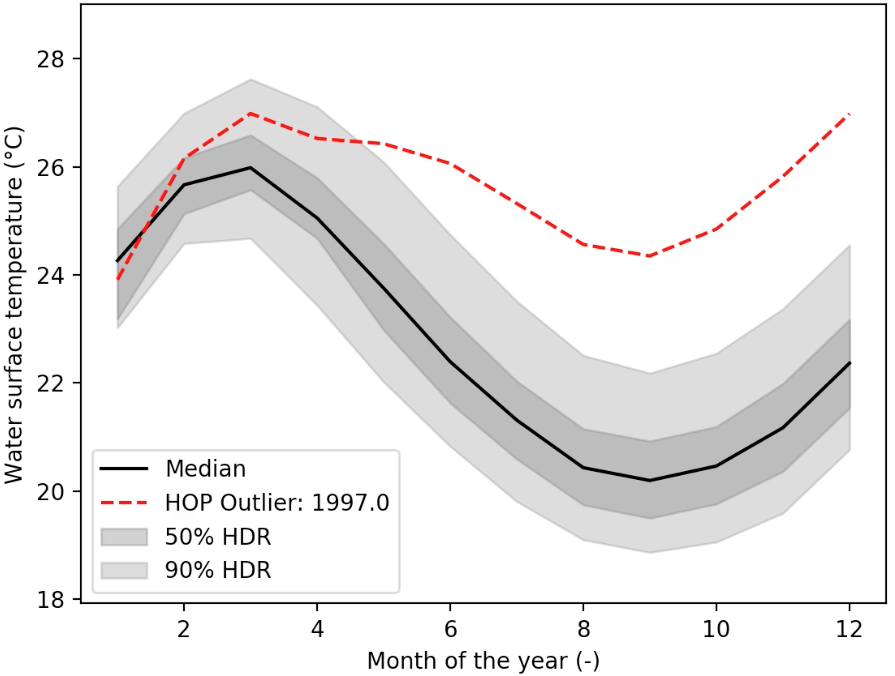}} 
\caption{Functional-HOPs. \textbf{a} close to median realization. \textbf{b} outlier realization. An animated version of this figure is available as supplementary material (\hyperref[S1]{S1}).}
\label{fig:f-hops}
\end{figure*}

\begin{figure}[!h]
\centering
\includegraphics[width=0.8\linewidth,keepaspectratio]{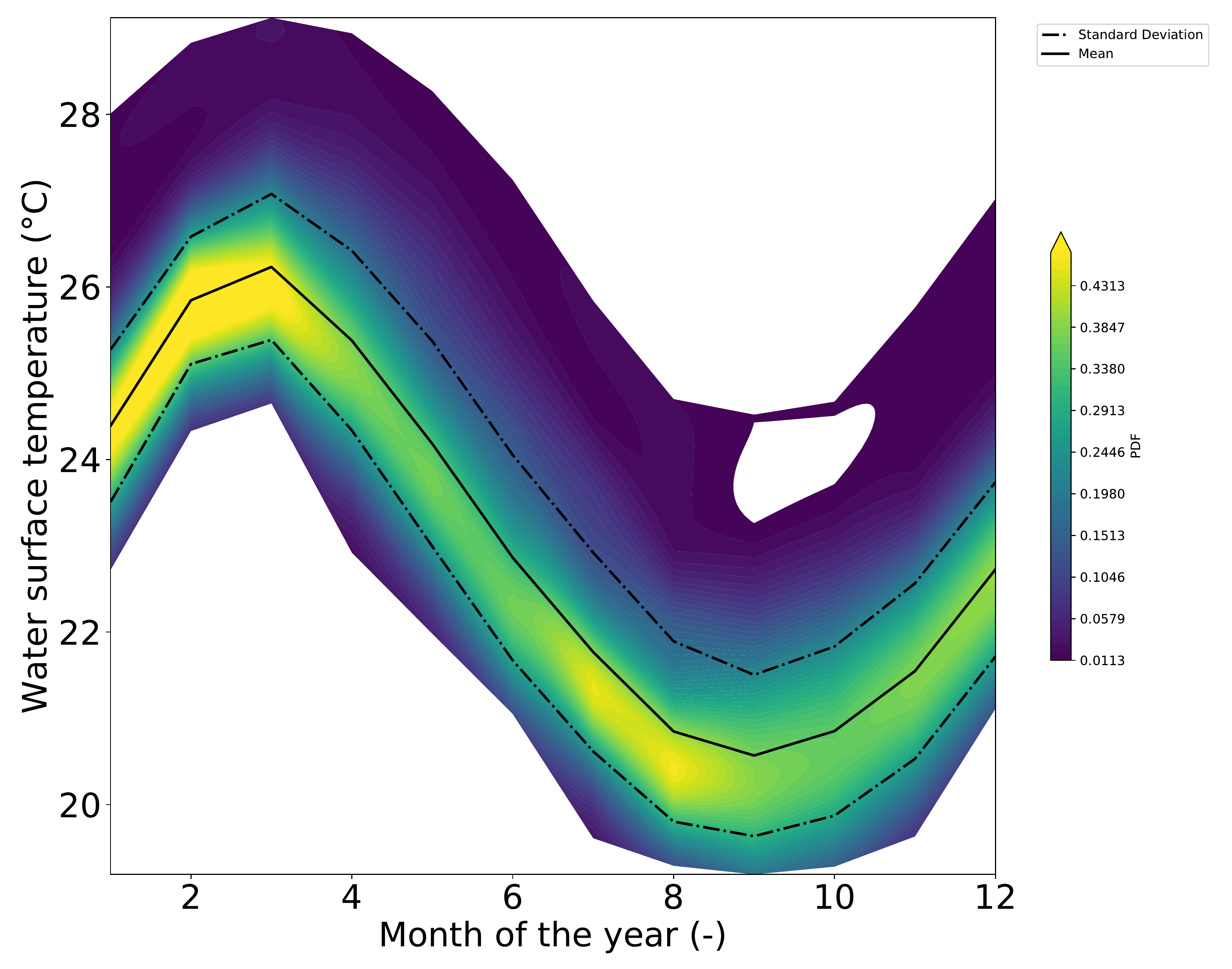}
\caption{Functional PDF of monthly sea surface temperature. Moments are computed with respect to the 58 realizations per month of the year.}
\label{fig:pdf}
\end{figure}

\begin{figure*}[!h]               
\centering
\subfigure[]{
\includegraphics[width=0.45\linewidth,height=\textheight,keepaspectratio]{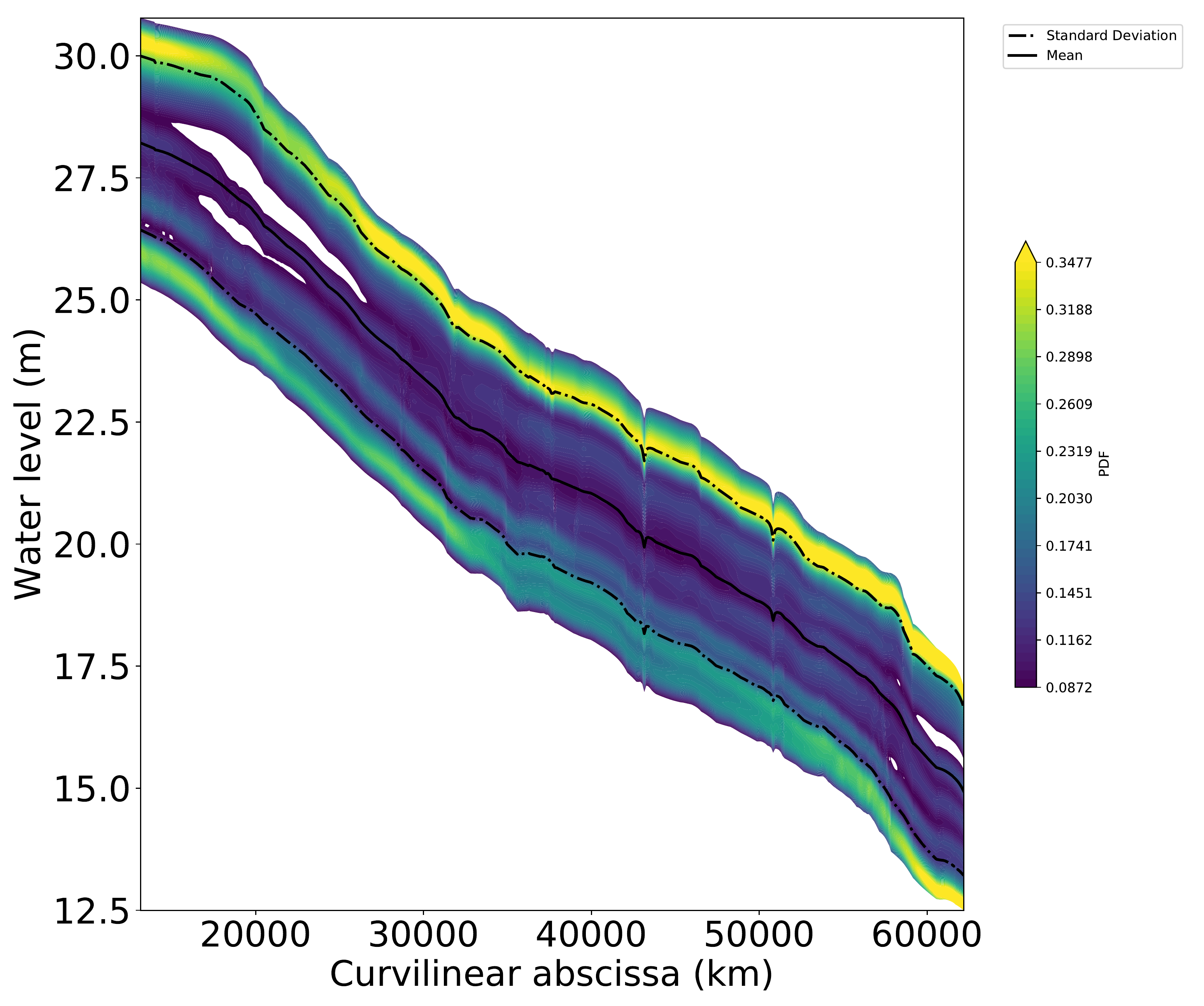}}
\subfigure[]{
\includegraphics[width=0.45\linewidth,height=\textheight,keepaspectratio]{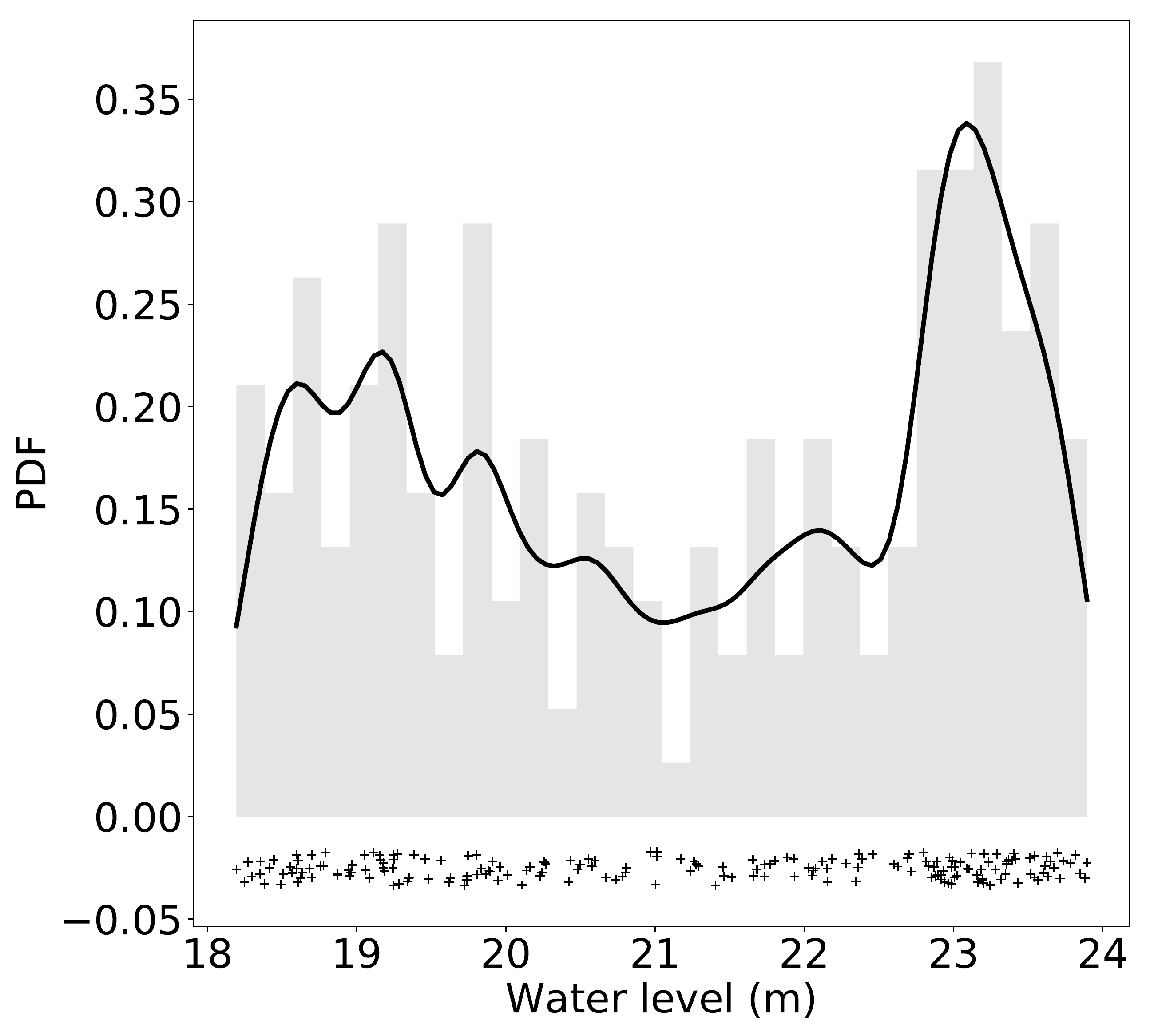}} 
\caption{Functional PDF of the water elevation, \textbf{a} along the 463 nodes along the river reach; \textbf{b} at Marmande station (36~km) revealing a bimodal PDF.}
\label{fig:pdf_MASCARET}
\end{figure*}

The HOPs dynamic visualization can be applied to functional outputs~\citep{Kale2018} as in~\cref{fig:f-hops} animating successive realizations of the data (for the El Ni\~no dataset), it is noted f-HOPs. When combined to HDR criteria, each realization is discriminated taking into account the functional characteristics of the output and outliers are easily detected. This solution is complementary to the classical PDF plot shown in~\cref{fig:pdf} that displays the probability of monthly sea surface temperature. The median and standard deviation curves are shown but these statistics are computed point by point independently and the outliers cannot be represented. Moreover, f-HOPs does not allow to exhibit multiple modes statistics that are shown by the PDF at a given location. \Cref{fig:pdf_MASCARET} shows both the functional PDF and a PDF at a precise location of the Hydrodynamics dataset. It can be noted that there are multiple modes.


\subsection{Sonification of Functional Outputs' Statistics}
\label{subsec:sound}

When the number of realizations in the dataset is limited, a static or an animated visualization, using f-HOPs and HDR metrics,  allow to depict the most significant characteristics of the ensemble members. When this number increases, the animation time increases and the analysis becomes harder. Sonification comes as an alternative for meaningful analysis of the dataset: it conveniently allows to draw the attention on specific realization discriminated by the HDR metric. We propose to compute the $L_2$-norm between each realization and a reference realization within the reduced modal space.  The reference realization is here chosen as the median and is mapped into a base sound. The frequency associated with each realization increases, and the sound becomes higher as the distance between each realization and the reference increases. Evidently, the reference can differ from the median. 
\citep{Alexander2014} presents that sonification allows for well-informed understanding of a large dataset and that practitioners usually develop a physics-dependent vocabulary that is adapted to specific need as in~\citep{Hughes2003} where sound is used to discriminate the types of gravitational waves. In this context, sonification serves data exploration. It  can also be an alert system: f-HOPs augmented with HDR metric sonification allows to get a fairly monotonous sound for realizations that are close to the median, while outliers are clearly spotted as proposed in~\hyperref[S1]{S1} for the El Ni\~no dataset.

\section{Uncertainty Visualization of Large Number of Input Variables}
\label{sec:input}

\begin{figure}[!h]
\centering
\includegraphics[width=0.8\linewidth,keepaspectratio]{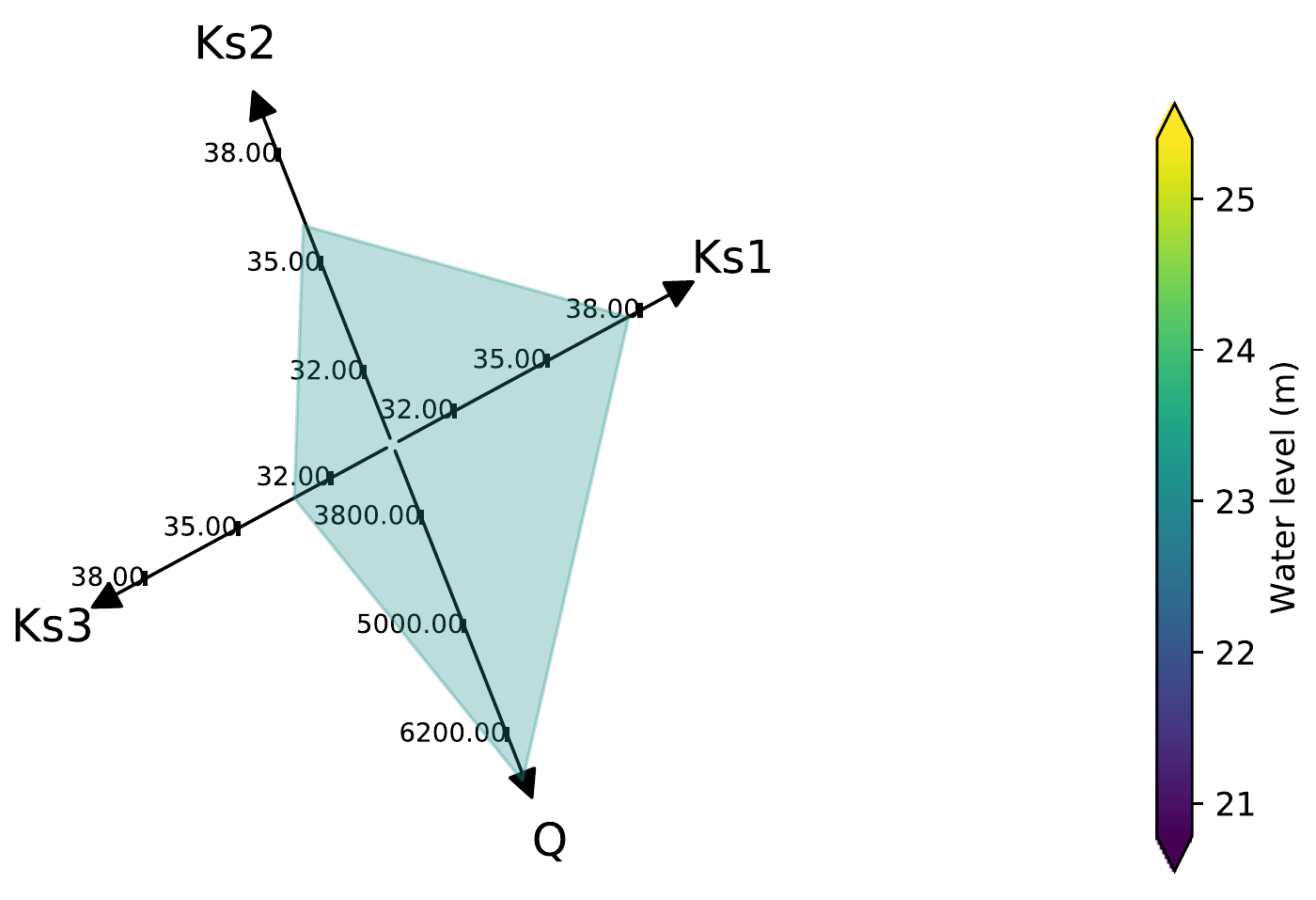}
\caption{Kiviat plot on the Hydrodynamics dataset: 2-dimensional Kiviat plane represents one sample of the data set. Colour maps the water level QoI value at Marmande (one location and one time).}
\label{fig:Kiviat_color}
\end{figure}

In the previous section, the focus was made on visualizing the QoI, not taking into account the relation between inputs and outputs. As visualizing the input alone, especially in case of small input dimension, is classical, the dual representation of input and output data remains challenging especially for large dimension functional outputs. The response surface plot is adapted when the input space is 2D or 3D. When the input space dimensions further increases, other solutions should be preferred such as parallel coordinates plot~\cite{Inselberg1985} or 3D-\emph{Kiviat} proposed by~\citep{Hackstadt1994}.

\begin{figure*}[!h]               
\centering
\subfigure[Stacking: QoI - Colouring: HDR]{
\includegraphics[width=0.47\linewidth,height=\textheight,keepaspectratio]{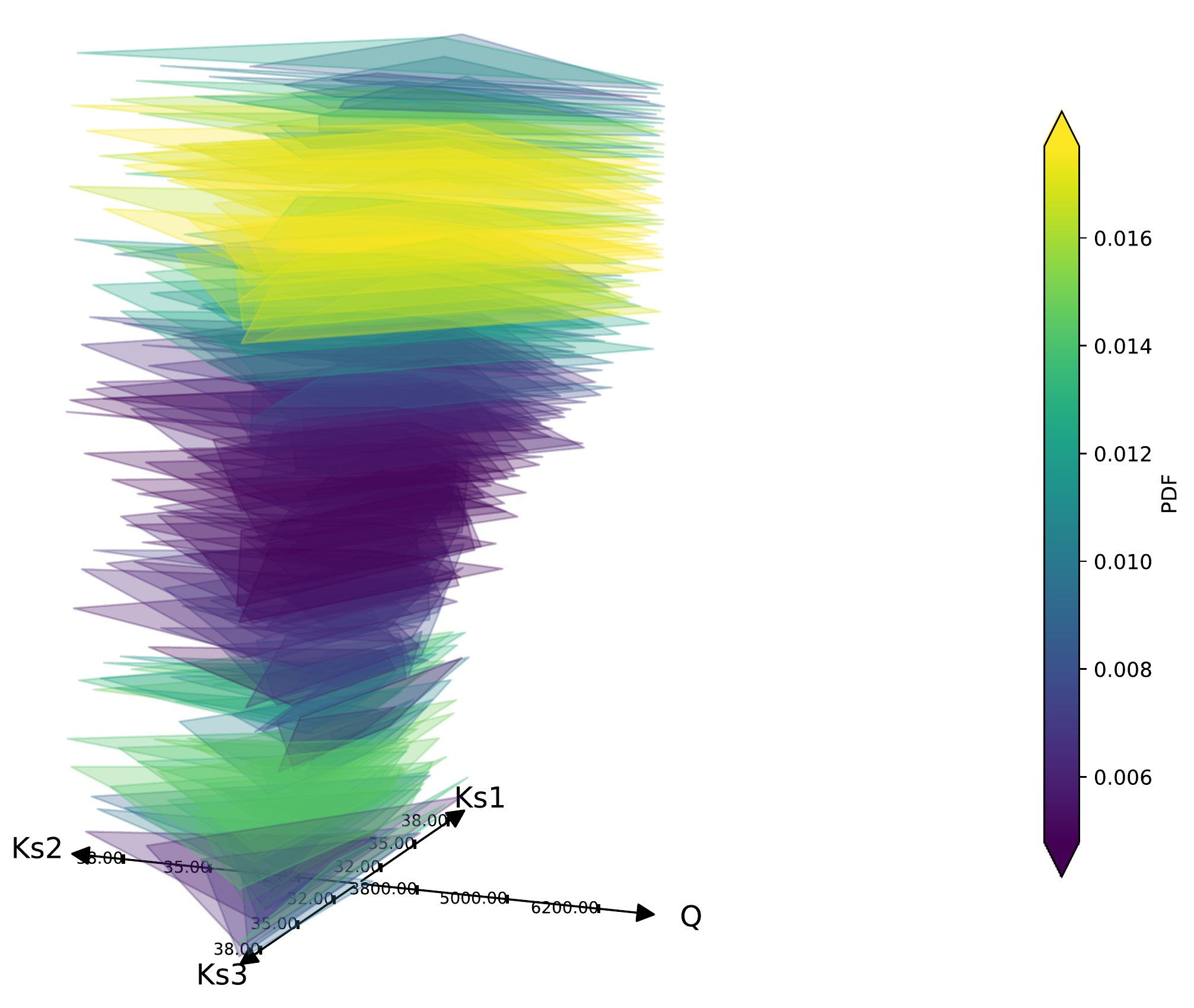}}
 ~       
\subfigure[Stacking: HDR - Colouring: QoI]{
\includegraphics[width=0.47\linewidth,height=\textheight,keepaspectratio]{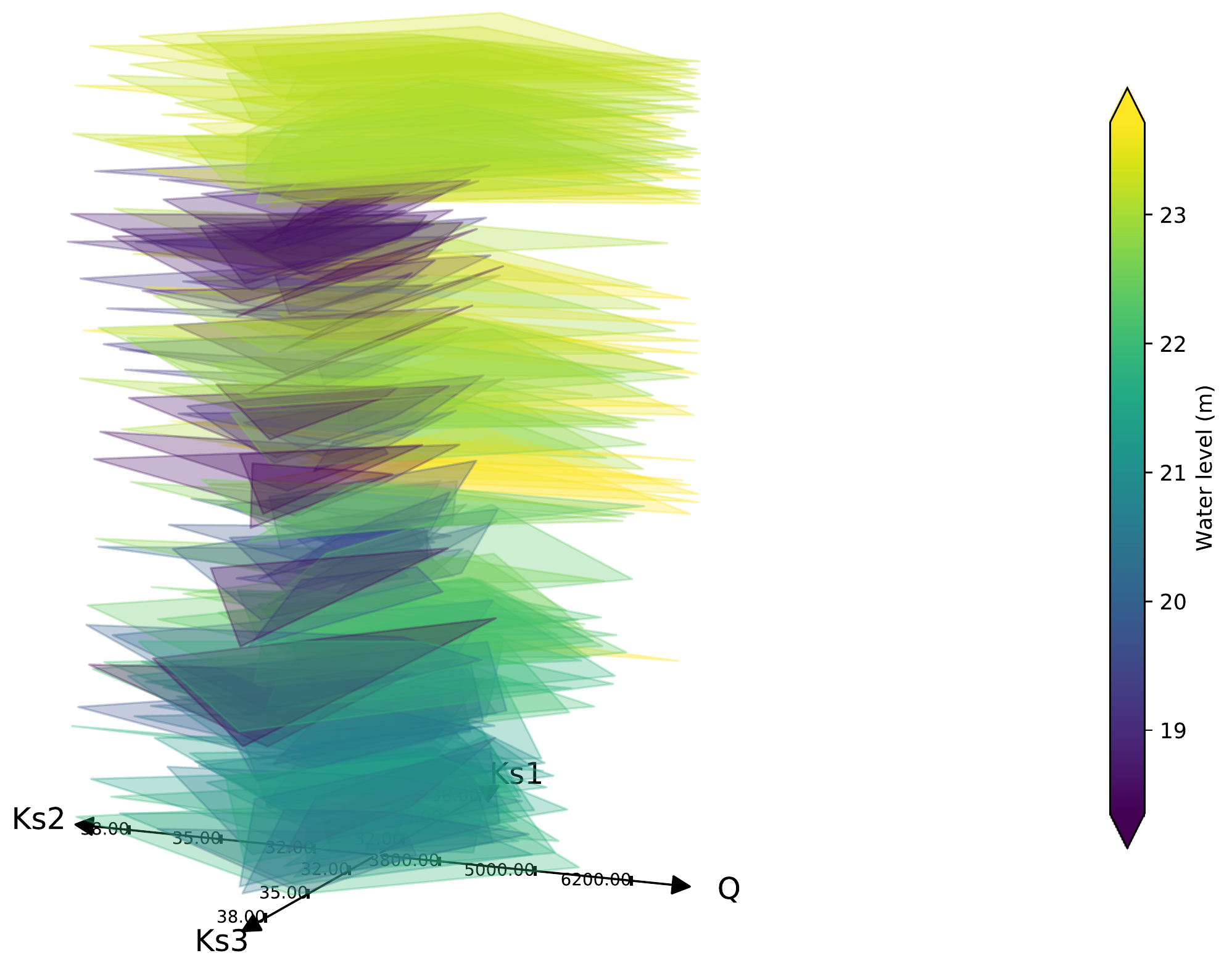}}

\subfigure[Stacking: $Q$ - Colouring: HDR]{
\includegraphics[width=0.47\linewidth,height=\textheight,keepaspectratio]{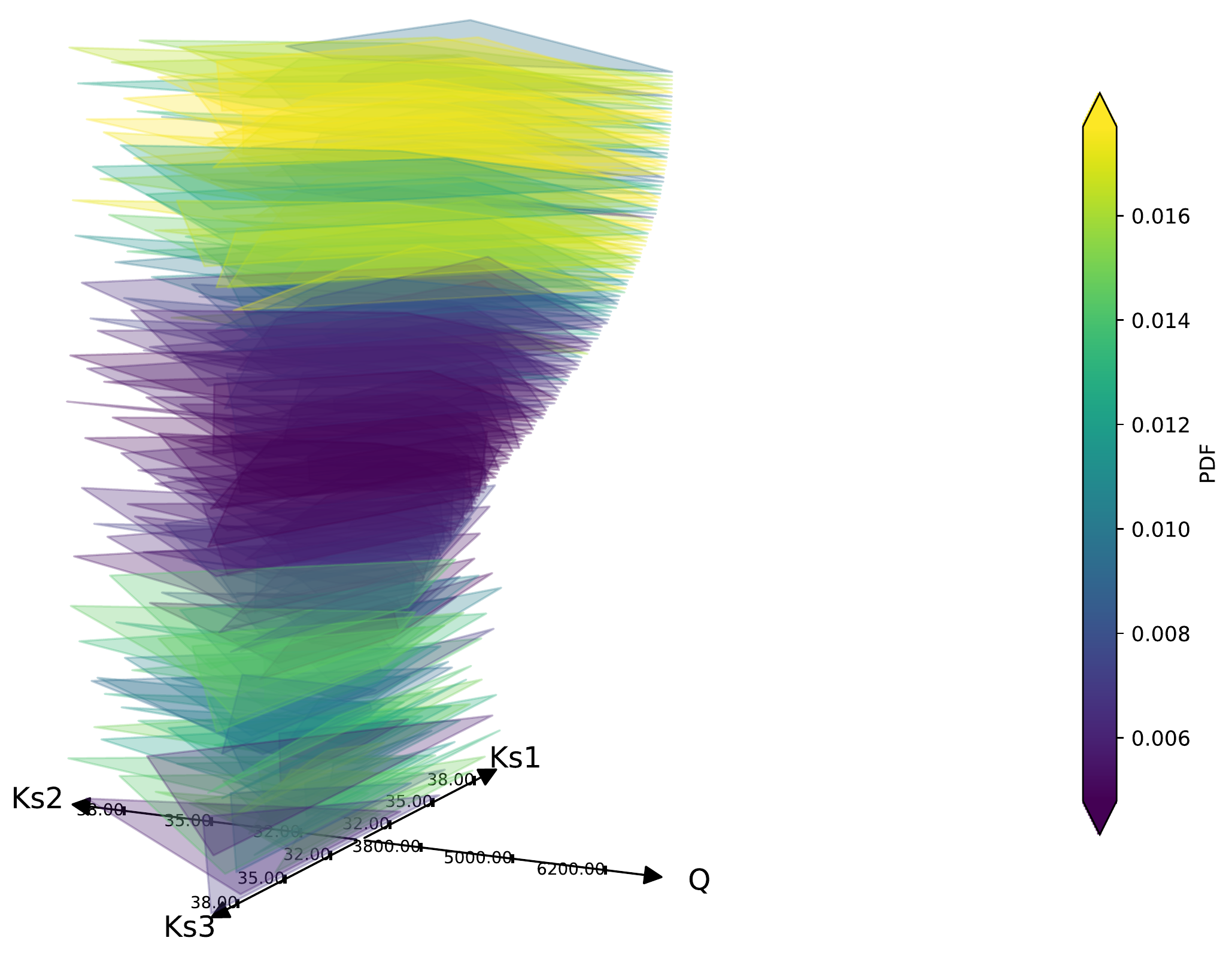}}
~
\subfigure[Stacking: QoI - Colouring: QoI]{
\includegraphics[width=0.42\linewidth,height=\textheight,keepaspectratio]{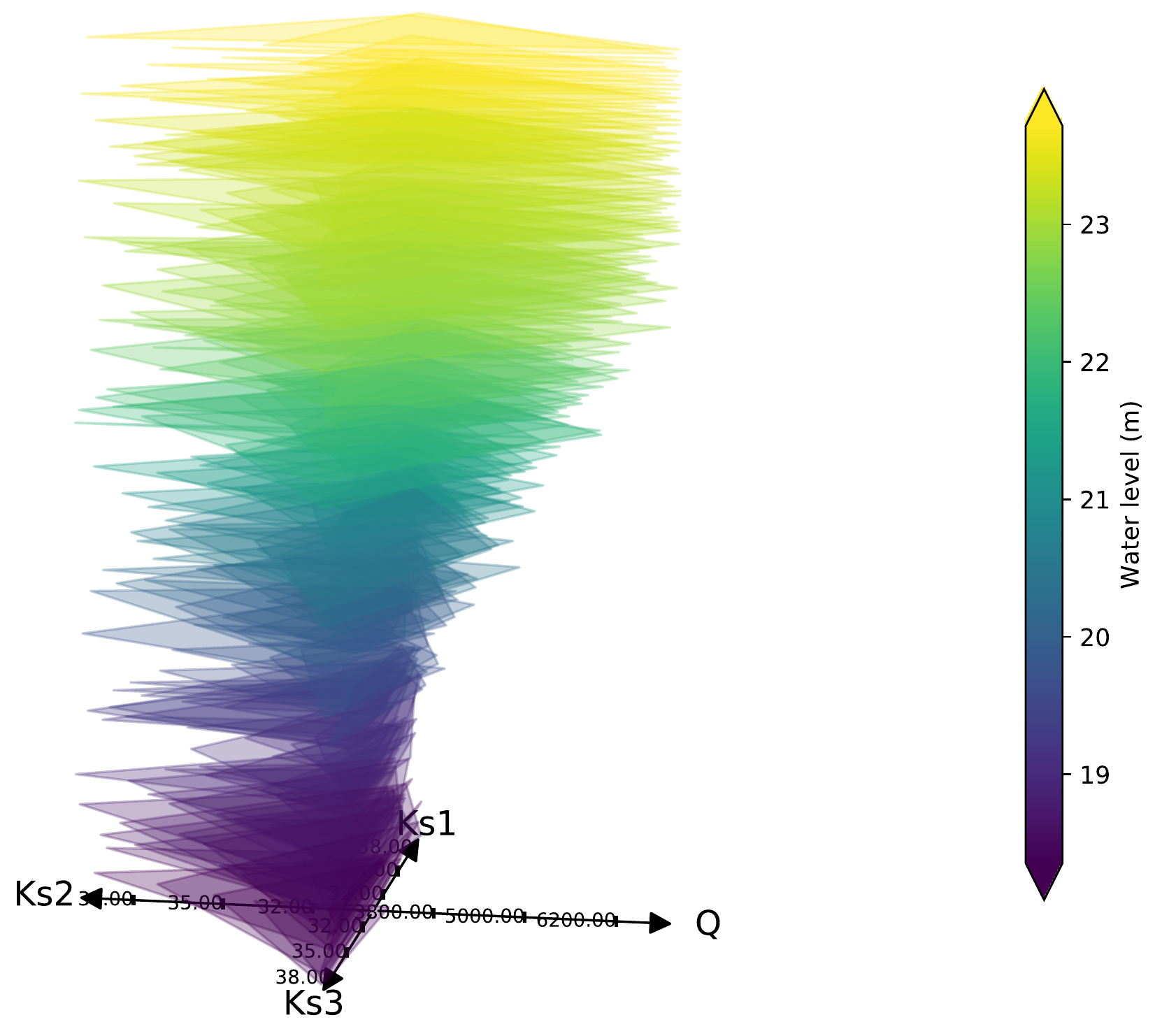}}
\caption{Kiviat plot on the Hydrodynamics dataset: comparison of different stacking orders and colour map strategies. \textbf{a} samples stacked by QoI and coloured by HDR. \textbf{b} samples stacked by HDR and coloured by QoI. \textbf{c} samples stacked by $Q$ and coloured by HDR. \textbf{d samples} stacked by QoI and coloured by QoI. An animated version of \textbf{a} is available (\hyperref[S2]{S2}).}
\label{fig:Kiviat_order}
\end{figure*}

In this paper, we propose to adapt the 3D-\emph{Kiviat} plot to the visualization of both input and QoI spaces. Each plane of the Kiviat represents a realization within the dataset with as many directions as the input dimensions as shown in~\cref{fig:Kiviat_color} for the hydrodynamics dataset. The input variables here correspond to the friction coefficients ($Ks1$, $Ks2$, $Ks3$) and the constant inflow $Q$; the output variable is the water level at Marmande, it is colour-coded onto the Kiviat plane. For the 3D-Kiviat, planes are stacked into a 3D object with respect to the QoI (scalar or functional) related value that is colour-coded. It should be noted that each plane is filled with only one colour to preserve readability. The benefit of 3D-Kiviat stands in the choice of both the stacking and the colouring strategies as shown in~\cref{fig:Kiviat_order} for the hydrodynamics dataset. Additionally, the 3D-Kiviat can be augmented with sound---as described in~\cref{subsec:sound}.

When representing functional output data, different stacking and colouring strategies allow to highlight different information in the dataset. Four choices of stacking and colouring are illustrated in~\cref{fig:Kiviat_order}; the stacking and colouring choices are indicated in the legend, they are achieved with respect to the QoI at a given location and time (noted $QoI$), with respect to the HDR metric (noted $HDR$) or with respect to one of the input variables. In~\cref{fig:Kiviat_order}(a), stacking is done with respect to the QoI at a given location and time while the colouring is done with respect to the difference to the median realization computed with the HDR metric. This allows to get a sense of the spatial PDF represented in~\cref{fig:pdf_MASCARET} augmented with the input parameter mapping. Another possibility in~\cref{fig:Kiviat_order}(b) consists in stacking with respect to the HDR metric and colouring with respect to the QoI value. In~\cref{fig:Kiviat_order}(c), stacking is done with respect to one of the input variable (in the present case $Q$) and the colouring is done with respect to the HDR metrics. Finally, \cref{fig:Kiviat_order}(d) displays both stacking and colouring with respect to the QoI value. It should be noted that an animated version of~\cref{fig:Kiviat_order}(a) is provided in supplementary material~\cref{S2} with sounding for the HDR metric that represents the distance to the median realization and allows for a convenient analysis of the data set especially when its dimension increases.

From~\cref{fig:Kiviat_order}(a, c, d), the impact of $Q$ on the water level is easily readable; water level increases with $Q$. High water level values are also obtained for low $Ks3$ values while other parameters seem to have no significant impact on the QoI. $Ks1$ and $Ks2$ have barely any impact on the QoI. The manipulation of the animated 3D-Kiviat (\cref{S2}) is even more adapted to data analysis. The coloured HDR in~\cref{fig:Kiviat_order}(a, c) indicates how each realization differs from the median realization. It appears that stacking for colouring with respect to QoI or HDR serves different purposes. Ordering by QoI allows to discriminate which input lead to specific QoI value while ordering by HDR illustrated the dispersion of the dataset with respect to a reference realization. Sounding is a supplementary way to emphasis the information, especially for large datasets.

3D-Kiviat comes as a complementary tool to classical sensitivity analysis criteria such as \emph{Sobol'} indices~\citep{Saltelli2007}. When computed on a large dataset (100 000 members) for the hydrodynamics example and averaged over space, these indices show that most of the variance of the water level is explained by $Q$ with $S_Q=0.98$. A small part of the variance is explained by $Ks3$ as $S_{Ks3} = 0.14$ and even smaller by $Ks1$ and $Ks2$ with $S_{Ks1}=0.01$ and $S_{Ks2} = 0.07$.

While the Sobol' indices quantify the importance of each parameter on the QoI's variance, they do not indicate the nature of these contributions. Indeed, it is stated that $Q$ has a significant impact but from~\cref{fig:Kiviat_order}(a, c, d), we can also add that the contribution is monotonous. Additionally, the weak impact of $Ks1$ and $Ks2$ is confirmed by the lack of a pattern in the 3D-Kiviat along these axes. Finally, while the $Ks3$ Sobol index is weak, the 3D-Kiviat indicates that small $Ks3$ values lead to high-water level values.

\begin{figure}[!h]
\centering
\includegraphics[width=0.6\linewidth,keepaspectratio]{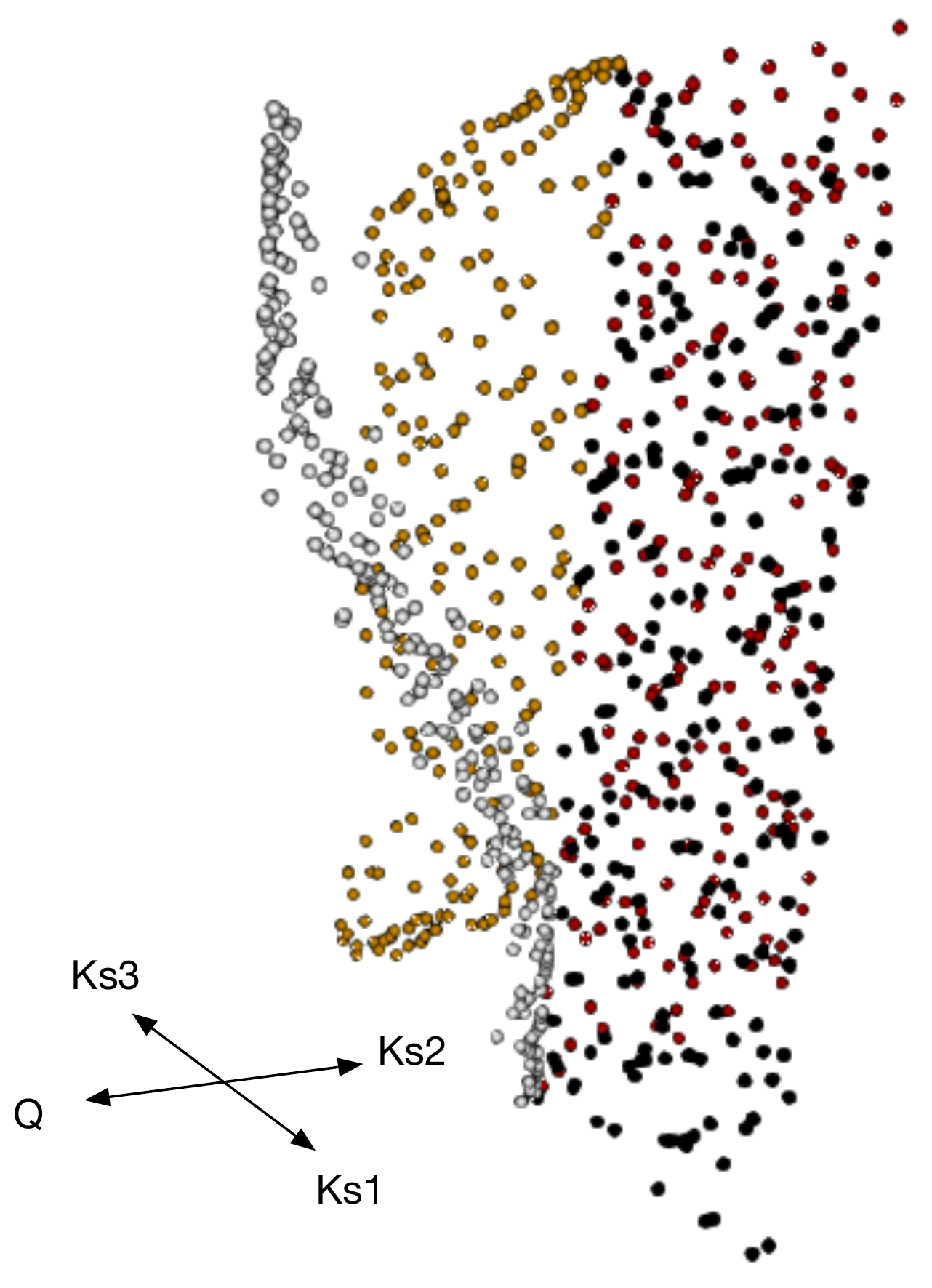}
\caption{3-dimensional Kiviat in point representation stacking with respect to the QoI and colouring with respect to the input parameter. An animated version of this figure is available with stacking with respect to the QoI and colouring with respect to the HDR metric in~\hyperref[S3]{S3} and with respect to the input parameter in ~\hyperref[S4]{S4}.}
\label{fig:mesh}
\end{figure}

It appears that 3D-Kiviat plot becomes difficult to manipulate when the dimension of the input parameter increases ($>10$). To overcome this caveat, we have designed a method to generate a surface mesh of the Kiviat plot as shown in~\cref{fig:mesh}. This allows the use of regular CAD viewers and thus facilitates the manipulation of the 3D object as well as the comprehension of the data structure. The construction is based on a vertex-vertex representation using quadrilateral elements. An animated version of~\cref{fig:mesh} is provided in~\cref{S3} with stacking with respect to the QoI value, colouring with respect to the HDR metric, then colouring with respect to the input parameters in~\cref{S4}. 


\begin{figure}[!h]
\centering
\includegraphics[width=0.8\linewidth,keepaspectratio]{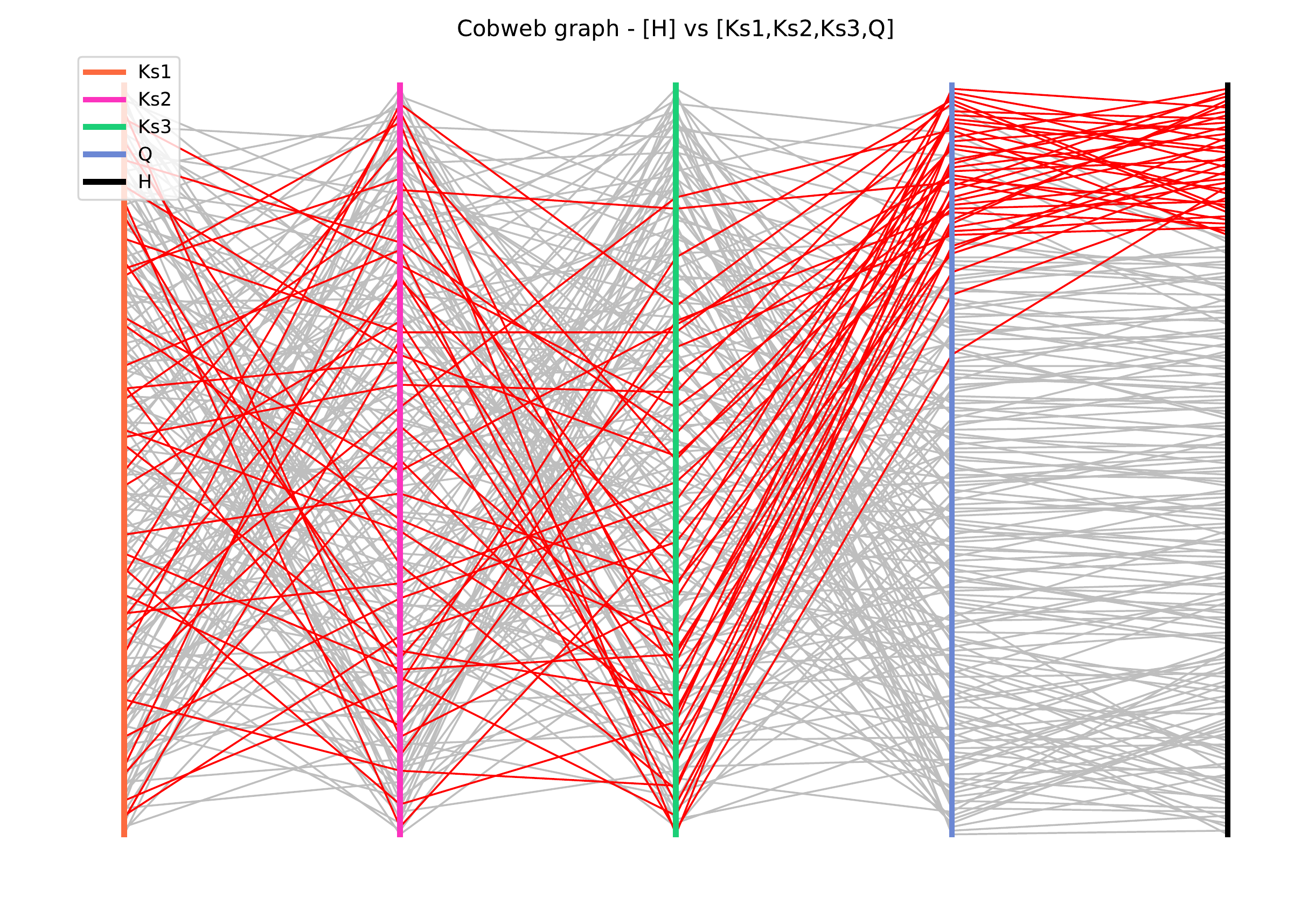}
\caption{Parallel coordinates plot for the hydrodynamics dataset with 80\% of the lowest values of QoI filtered out (in grey) and high values in red.}
\label{fig:cobweb-Kiviat}
\end{figure}

The analysis from 3D-Kiviat is similar to that drawn from the parallel coordinates plot in~\Cref{fig:cobweb-Kiviat}. The latter consists of $n+1$ parallel axes, with $n=4$ the number of input parameters for the hydrodynamics dataset. The last axis is dedicated to the output value, here water level at Marmande. Each grey line corresponds to one realization in the dataset; the red lines are discriminated for high-water level values resulting from high $Q$ and small $Ks3$, independently of $Ks1$ and $Ks2$. It should be noted that parallel coordinate plot is not adapted to simultaneous representation of QoI and HDR metrics contrary to 3D-Kiviat. Finally, the linear relationship between water level and $Q$ that was clear on the 3D-Kiviat in not readable on the parallel coordinate plot that is more adapted to clustering with the possibility of advanced strategies proposed by~\citep{Raidou2016}.

\begin{figure}[h]
\centering
\includegraphics[width=0.6\linewidth,keepaspectratio]{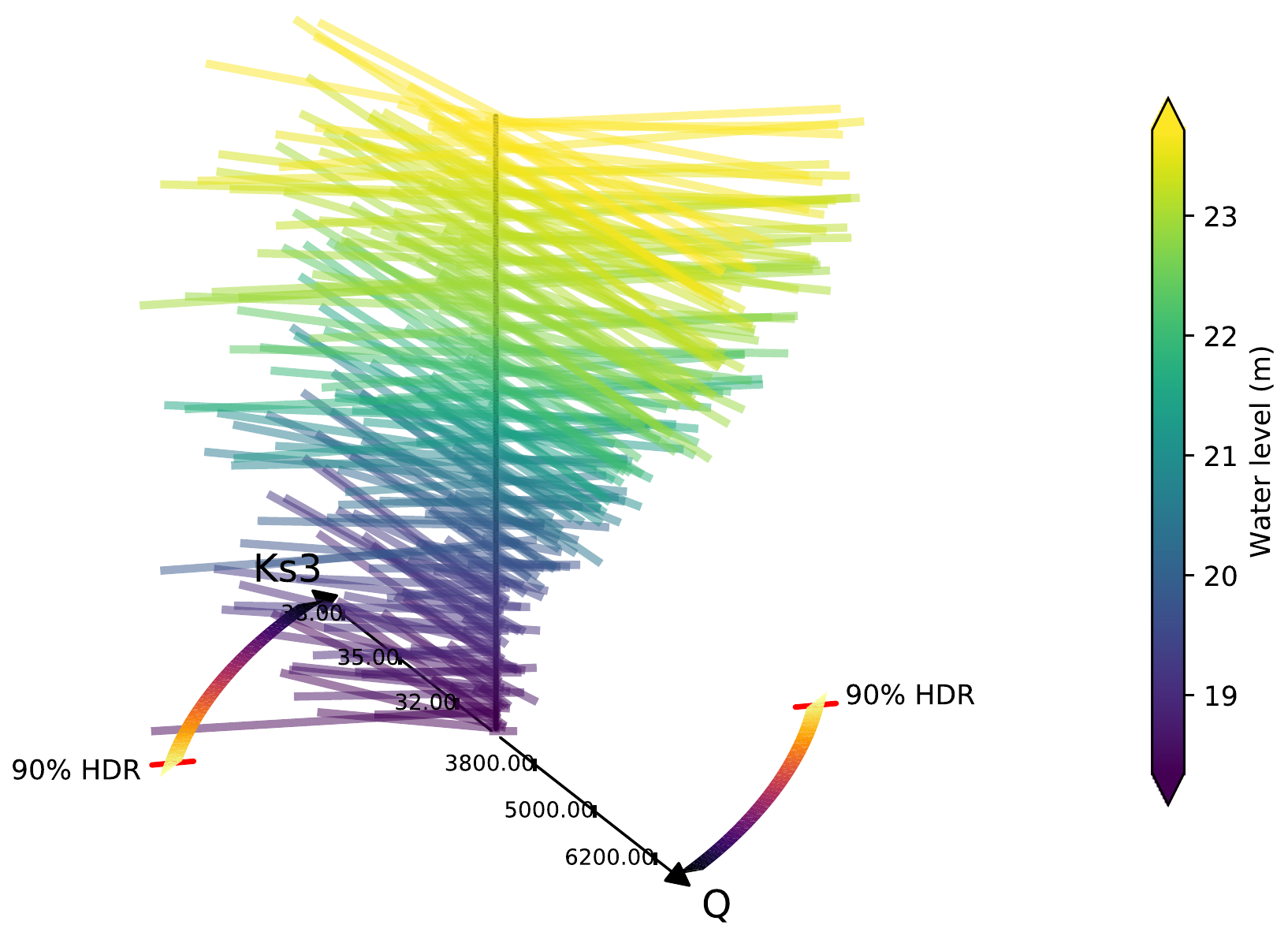}
\caption{Tree plot for the hydrodynamics dataset considering the most significant parameters $Ks3$ and $Q$. Here the QoI (water level at Marmande) is represented by stacked and coloured segments and HDR metric is represented by the azimutal component.}
\label{fig:tree}
\end{figure}

The specific case of a 2D input space is treated with a \emph{Tree plot} solution where Kiviat coloured planes are replaced by coloured segments that are stacked and coloured regarding to QoI related value. The vertical stacking and colouring are achieved with respect to the QoI value. The HDR metric is encoded as an azimutal component: the angle is null if the realization corresponds to the median and the angle increases as the realization differs to the median. \Cref{fig:tree} displays a tree plot for the hydrodynamics dataset where $Ks1$ and $Ks2$ are not accounted for (since they were previously shown to have barely any impact on the QoI) and the QoI is the water level at Marmande. Here again, stacking, colouring, angle and eventually sounding strategies can be adapted to convey information on a dataset and efficiently enhance meaningful information.


\section{Conclusion and Discussion}
\label{sec:ccl}

This work proposes a systematic way to look at uncertainties when dealing with high-dimensional environment. As uncertainty quantification related analysis are moving toward high dimensionality, we believe that our tool could help these analyses.

In this work, a new method to visualize input and output uncertainties  was presented. The concept of Hypothetical Outcomes Plot is taken a step further and applied to functional data (f-HOPs) considering the HDR boxplot information that estimates a distance to a reference realization (here the median). HDR metrics allow to reduce the output space to a limited number of principal components and compute, in this reduced space, a distance between dataset realizations taking into account temporal or spatially discretization. This metric allows to discriminate the realizations statistically instead of only regarding the QoI value at a specific location and time. f-HOPs provides an animated version of members within a dataset that can be augmented with data sonification that is efficient for outlier detection. Both input and output uncertainties are then visualized with a 3-dimensional version of the Kiviat where each realization is represented by a plane surface. Realizations are stacked  along the vertical axis and colour-coded. The 3D-Kiviat can be augmented with data sonification. Stacking, colouring and sounding strategies are chosen with respect to the QoI value or the HDR metric to highlight some information within the dataset.  

These visualization solutions were applied to two functional datasets: \emph{(i)}~El Niño and \emph{(ii)}~Hydrodynamics. The first one is a commonly used dataset that only provides temporal outputs, the second dataset presents an input-output relation with a large dimension output space and the input space larger than the classically represented 3D space. For both datasets, f-HOPs and 3D-Kiviat have proven to be efficient at representing the nature of the dataset and highlighting important impact factors. Moreover, visualization of the 4-dimensional input space was eased by the Kiviat solution that was conveniently adapted to CAD objects for larger dimensions. 

Perspective for this work stands in the adaptation of stacking, colouring and sounding to specific purposes. For instance, for extreme events detection, the HDR metrics can be computed with respect to the mean or the median, but it could also be computed with respect to high quantiles.
The reference can also represent additional information to the dataset. For instance, in the context of data assimilation, it can stand for observation data that are usually distributed in space and time, in order to compute the innovation vector or to achieve observation quality control steps.

Our strategy finally can be applied to operational context with ensemble integration. Straightforward applications are for instance real-time weather and flood forecasting as well as structure failure risk assessment and outliers detection are at stake. In this context, our strategy would also be applied to larger dimension input and output spaces. The stacking strategy allows to create a 3D object that can easily be manipulated, even printed for deep analysis. Sounding is also an efficient way to draw the attention to an operator while the ensemble forecast is issued in case threshold exceed occurs. Finally, the complete solution with stacking, colouring and sounding offers social and human perspectives for statistical analysis of datasets by people with disabilities for which written information is not handy.  

\bigskip
\begin{center}
{\large\bf SUPPLEMENTARY MATERIAL}
\end{center}

\begin{description}


\item[S1:] Sounding Functional HOPs on the El Niño dataset. (movie file with audio) \label{S1}

\item[S2:] Sounding 3-dimensional Kiviat on the Hydrodynamics dataset. (movie file with audio) \label{S2}

\item[S3:] Mesh animation on the Hydrodynamics dataset, colouring with respect to HDR metric. (movie file) \label{S3}

\item[S4:] Mesh animation on the Hydrodynamics dataset, colouring with respect to input parameters. (movie file) \label{S4}

\item[S5:] open-source (MIT) Python script used to do HDR boxplot, Kiviat and data sonification. (python file)

\item[batman:] open-source (CECILL-B) Python package to perform uncertainty quantification on any experiment~\citep{Roy2018}. The package contains all functions and datasets used as examples in the article. (\href{https://gitlab.com/cerfacs/batman}{https://gitlab.com/cerfacs/batman})

\end{description}

\bibliographystyle{plainnat}
\bibliography{papiervisu.bib}
\end{document}